\documentclass[aps,prd,twocolumn,preprintnumbers,superscriptaddress,floatfix]{revtex4}

\usepackage{multirow, graphicx,amssymb,url,mathrsfs,amsmath}
\usepackage{eucal,wrapfig,boxedminipage,setspace,subfigure}
\usepackage{amsxtra,amstext,latexsym,dsfont}

\def\IR{{\hbox{{\rm I}\kern-.2em\hbox{\rm R}}}}
\def\IB{{\hbox{{\rm I}\kern-.2em\hbox{\rm B}}}}
\def\IN{{\hbox{{\rm I}\kern-.2em\hbox{\rm N}}}}
\def\IC{\,\,{\hbox{{\rm I}\kern-.59em\hbox{\bf C}}}}
\def\IZ{{\hbox{{\rm Z}\kern-.4em\hbox{\rm Z}}}}
\def\IP{{\hbox{{\rm I}\kern-.2em\hbox{\rm P}}}}
\def\IH{{\hbox{{\rm I}\kern-.4em\hbox{\rm H}}}}
\def\ID{{\hbox{{\rm I}\kern-.2em\hbox{\rm D}}}}

\newcommand{\beq}{\begin{equation}}
\newcommand{\eeq}{\end{equation}}
\newcommand{\bea}{\begin{eqnarray}}
\newcommand{\eea}{\end{eqnarray}}

\begin{document}

\voffset 1cm

\newcommand\sect[1]{\emph{#1}---}

\title{Running anomalous dimensions in holographic QCD:\\ from the proton to the sexaquark}

\author{Nick Evans}
\affiliation{ STAG Research Centre \&  Physics and Astronomy, University of
Southampton, Southampton, SO17 1BJ, UK}

\author{Matthew Ward}
\affiliation{ STAG Research Centre \&  Physics and Astronomy, University of
Southampton, Southampton, SO17 1BJ, UK}

\begin{abstract}
In holographic models of QCD, the running of the anomalous dimension of the quark bilinear operator leads to chiral symmetry breaking when $\gamma=1$ and the Breitenlohner-Freedman bound is violated. In that case, the running drives the sigma meson mass tachyonic inducing the chiral symmetry breaking. Here we include the running anomalous dimension in the computation of the spectrum of bound states associated with other operators made of light quarks, such as the nucleon and exotic sexaquark states. We show that including the one loop gauge theory running can have substantial effects on the predictions.  For example, the nucleon mass to rho mass ratio is improved and lies much closer to the observed value. A similar result is obtained for the $\Lambda$ and $\Xi$ baryon when strange quarks are included. A $uuddss$ sexaquark state with a low enough mass to make it stable can be achieved, but this depends on the input assumptions about the running dimension.

\end{abstract}%

\maketitle

\newpage

\section{Introduction}\vspace{-0.5cm}

Holographic models \cite{Erlich:2005qh,DaRold:2005mxj} of QCD are remarkably simple, consisting of free scalars, vectors or fermions in a bulk Anti de-Sitter (AdS) space. The fields' bulk masses (in units of the AdS radius) are set by the holographic dictionary \cite{Witten:1998qj}, for example,  the scalar mass $m^2 = \Delta(\Delta-4)$ where $\Delta$ is the scaling dimension of the dual field theory operator. Simple AdS/QCD models, with the mass gap included via a hard wall, are already successful in predicting the light mesonic spectrum at the 15\% level \cite{Erlich:2005qh,DaRold:2005mxj}. The pattern of baryon masses is also well produced by treating them as spinors in AdS \cite{deTeramond:2005su} but the nucleon to rho meson mass ratio typically comes out $\sim$50\% too high. These hadrons arise from the lowest energy normalisable modes in the bulk. The justification for these models is provided by the Dirac-Born-Infeld (DBI) action of probe branes in top-down models, such as the Sakai-Sugimoto model \cite{Sakai:2004cn} or the D3/probe D7 system \cite{Karch:2002sh, Kruczenski:2003be, Erdmenger:2007cm}.

Top-down string constructions of holographic duals to chiral symmetry breaking \cite{Babington:2003vm, Kruczenski:2003uq} show that, upon deforming the AdS background (to describe field theories beyond ${\cal N}=4$ super Yang-Mills), the DBI action is altered in a way that provides a running anomalous dimension to the quark operators it describes \cite{Alvares:2012kr}. In bottom-up models of QCD, the running of the $\bar{q} q$ state is usually imposed by hand, either directly \cite{Alho:2013dka,Erdmenger:2020flu} or via a constructed scalar potential \cite{Jarvinen:2011qe}. In the bulk, the dual field's mass changes with radial position, becoming tachyonic (i.e. violating the Breitenlohner-Freedman (BF) bound \cite{Breitenlohner:1982jf}) in the infra-red. For a scalar, this instability sets in when the anomalous dimension of the quark condensate equals 1 or, equivalently, when the bulk mass falls to $-4$.  As such, the formation of a quark condensate offers a dynamical alternative to the hard wall mechanism of the original models.

The vev of the $\bar{q} q$ operator in QCD implies that the $\sigma$ meson (the scalar isospin singlet of the $SU(N)_L \times SU(N)_R$ chiral symmetry) has become tachyonic as a result of strong interactions. Given the substantial effect on the $\sigma$'s mass, one might imagine that including the running dimension of operators is crucial when calculating the mass of further states in the spectrum. However, this has scarcely been investigated beyond the scalar sector in AdS/QCD. One reason is that vector and axial-vector states are described by gauge fields with vanishing masses in the bulk, indicating the absence of any such running.

That being said, such effects may be important for spin-\textonehalf ~ baryons composed of $u$, $d$ and $s$ quarks, such as the proton, $\Lambda$ and $\Xi$ baryons. Indeed, mass predictions for these states have typically been overestimated in AdS/QCD, which we demonstrate by calculating the nucleon mass within a simple hard wall model. Following this initial estimate, we construct a basic holographic model based on  the D3/probe D7 system that incorporates the running anomalous dimension for both the $\bar{q} q$ operator and the baryons. We use the one loop results from QCD which, as we will show, replaces these runnings with a product of the anomalous dimensions for the individual quark fields that constitute the operators. In this model, we find that our estimate for the nucleon mass is brought down by over a third, aligning both it and the $\Lambda$ and $\Xi$  baryon masses much more closely with their measured values.

Following these results, we turn our attention to a proposed, but so far undetected, particle in the QCD spectrum, the sexaquark \cite{Farrar:2017eqq, Farrar:2022mih}. Modelling the tightly-bound $uuddss$ six-quark state as a scalar in the AdS bulk, we consider a range of possible runnings for the operator's anomalous dimension. One can obtain a wide range of answers depending on the choice of running but, interestingly, opting for the one loop result (extended into the non-perturbative regime) leads to a mass that falls within the proposed range for it to remain stable against decay into two $\Lambda$s. We discuss some other aspects of the sexaquark's holographic phenomenology also.

\begin{figure}[h!]
    \centering
    \includegraphics[width = 8cm]{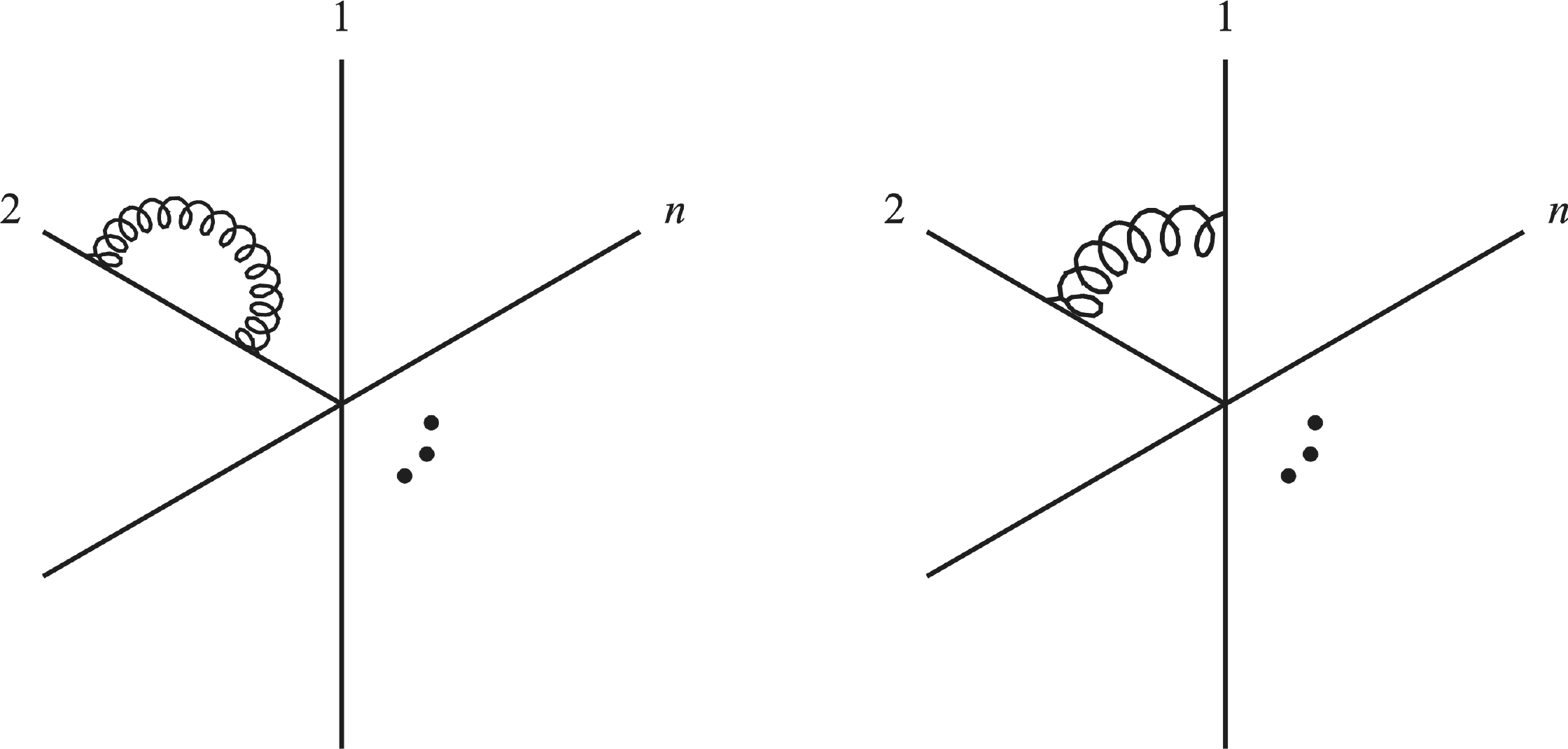}
    \caption{Diagrams at one loop order contributing to the anomalous dimension of a gauge invariant scalar operator with $n$ quark legs.}
    \label{fig:feynman_sexaquark}
\end{figure}

\section{Anomalous Dimensions at One Loop}
The crucial new ingredient we wish to explore in AdS/QCD models is the anomalous dimension of multi-quark operators. Let us therefore review the theory of anomalous dimensions in QCD at one loop level (covered in more detail in \cite{Grozin:2005yg}).

The dimension $d$ for a generic operator ${\cal O}=\tilde{\cal O} \mu^d$, where $\tilde{\cal O}$ is the dimensionless component of ${\cal O}$ at the scale $\mu$, can be expressed as
\begin{equation}
d = {1 \over {\cal O}} ~ \mu {d {\cal O} \over d \mu}
\end{equation}
Upon renormalisation by $Z_{\cal O}$, the anomalous dimension $\gamma$ is then given by
\begin{equation}
    \gamma_{\cal O} = - {1 \over Z_{\cal O}} ~\mu {d Z_{\cal O} \over d \mu} \label{eq:anon_dim}
\end{equation}

where we have included a minus sign such that $d = d_0 + \gamma$. Computing $Z_{\cal O}$ for a colour singlet gauge invariant operator comprised of $n$ quark fields requires one to consider two sets of Feynman diagrams (fig.\ref{fig:feynman_sexaquark}). Firstly, there is a factor $Z_\psi$ originating from the wave function renormalisation of the $n$ external legs, which diverges in the $\epsilon \rightarrow 0$ limit as per the dimensional regularization procedure
\begin{equation}
    Z_\psi = 1- C_2(R) ~\xi ~{\alpha \over 4 \pi} ~{1 \over \epsilon} \label{eq:wavefunc_renorm}
\end{equation}
where $C_2(R)$ is the quadratic Casimir element. We also define the square of the gauge coupling $\alpha =  g^2/4\pi$ and the gauge fixing parameter $\xi$, which specifies the form of the gluon propagator
\begin{equation}
    D^0_{\mu \nu}(p^2) = {-i \over p^2} \left(g_{\mu \nu} - (1-\xi) {p^\mu p^\mu \over p^2} \right) 
\end{equation}

Secondly, one must consider diagrams describing the exchange of gluons between any two external lines as permitted by the colour symmetry of the scalar operator. We have
\begin{equation}
    Z_V = 1 + (3 + \xi) ~{\alpha \over 4 \pi} ~{1 \over \epsilon}
\end{equation}
which, combined with \eqref{eq:wavefunc_renorm}, defines the renormalisation factor for the $n$ quark operator
\begin{equation}
    Z_{{\cal O}n} =   \left(1 + C_n(3 + \xi) ~{\alpha \over 4 \pi} ~{1 \over \epsilon}\right)Z_\psi^{n/2}
\end{equation}
where $C_n$ is a combinatoric colour factor discernible from the permitted diagrams. To ensure that our operator remains gauge invariant, we impose that $Z_{{\cal O}n}$ is independent of $\xi$, thereby fixing $C_n=n C_2(R)/2$. This gives
\begin{equation}
     Z_{{\cal O}n} = 1 + {3 n \over 2}  {C_2(R)}   ~{\alpha \over 4 \pi} ~{1 \over \epsilon}
\end{equation}

In the fundamental representation with $N_c = 3$ colours, we find that $C_2(F) = (N_c^2-1)/2N_c = 4/3$. Combining this with the result of \eqref{eq:anon_dim}, we can derive a general expression for the anomalous dimension of an $n$ quark vertex
\begin{equation}
    \gamma_{{\cal O}n} (\mu) = - n ~{\alpha (\mu) \over \pi} \label{eq:op_dim_general}
\end{equation}
As \eqref{eq:op_dim_general} shows, the anomalous dimension is directly proportional to the number of quark legs. Beyond one loop, this factorization is expected to break down and operators will garner their own distinct running. In the following sections, we will use this one loop result in AdS/QCD models which we have extended into the non-perturbative regime. Although results at two loop and beyond will begin to distinguish operators with different colour wave functions, one would need the full non-perturbative running to correctly understand the resultant splittings.
\vspace{-0.7cm}

\section{Hard Wall Model}
The simplest AdS/QCD model \cite{DaRold:2005mxj,Erlich:2005qh} involves the study of bulk fields in $AdS_5$ space with metric
\begin{equation}
    ds^2 = r^2 dx_\mu dx^\mu + \frac{dr^2}{r^2} 
\end{equation}
where $x^\mu \in \mathbb{R}^{1,3}$ and $r$ is the radial ordinate of the curved AdS space. In the dual field theory, this radial direction corresponds to the renormalisation group (RG) scale, allowing us to insert a mass gap into QCD by placing a hard wall boundary at $r=1$.

A simple example of extracting dynamics from the hard wall model is to consider a $SU(N_f)$ vector field in $AdS_5$ space with the action
\begin{equation}
    S = \int d^5x \textrm{Tr} \sqrt{-g} F^2
\end{equation}

The equation of motion in the $A^r=0$ gauge where $A_\perp = A(r) e^{ikx}$ and $k^2=-M^2$ is
\begin{equation}
    \partial_r r^3 \partial_r A + {M^2 \over r}A = 0
\end{equation}
For large $r$, the solutions take the form $A = A_0 + \bar{q}\gamma^\mu q/r^2$, with the two constants playing the role of background flavour gauge fields and the vector current respectively. Solutions that asymptote to $A=0$ pick out the rho meson states. In units where the wall is at $r=1$, the $\rho$ meson ground state has bulk mass $M^2 = 5.9$. Throughout this paper, we will use the computation of this mass in a given model to set the QCD scale. For these calculations, we use $M_\rho = 770$ MeV.

\subsection{Prediction of the Nucleon Mass}
A spin-half baryon can be described by a bulk spinor field $\psi$ with mass $m_\psi$ \cite{deTeramond:2005su,Abt:2019tas}. One can then square the linear Dirac equation to find the Klein-Gordon equation for the two components projected by $\gamma^0$; the linear equation linking the form of the solutions for each. In $AdS_5$, the suitable Klein-Gordon equation is
\begin{equation}
    \left( \partial_r^2 + {6 \over r} \partial_r + {6 - m_\psi^2 + m_\psi \over r^2} + {M_B^2 \over r^4} \right) \psi = 0 \label{eq:fermion_eom}
\end{equation}
where $m_{\textrm{eff}}^2 = 6 -m_\psi^2 + m_\psi$ defines an effective mass. The holographic dictionary then allows one to express this mass as a function of the scaling dimension associated with the boundary operator which, for a fermion field in $AdS_5$, takes the form $m_{\textrm{eff}}^2 = \Delta (5-\Delta)$. Solving this equation for a three quark state (i.e. $\Delta=9/2$), the bulk fermion mass is given by $m_\psi=5/2$ and the solution to \eqref{eq:fermion_eom} takes the following form (as given in \cite{Abt:2019tas})
\begin{equation}
   \psi =  {J M_B \over 4} {1 \over r^{5-\Delta} }+ {O \over r^\Delta} 
\end{equation}
Assuming the solution approaches the hard wall smoothly and is normalised to unity, i.e. the boundary conditions $\psi(1) = 1$ and $\psi'(1) = 0$ are imposed, we can tune the $M_B$ mass such that $\psi$ vanishes in the UV (i.e. the source $J$ is set to zero). Relative to the rho mass, we find that $M_B = 2.2 M_\rho \approx 1.7$ GeV.

This prediction is likely too large because we have neglected to include the running dimension of the bound state operator. The dimension of each quark in $\bar{q}q$ has fallen to one at the hard wall in order to trigger chiral symmetry breaking. By \eqref{eq:op_dim_general}, the dimension of a three quark operator should therefore have fallen to $\Delta=3$ in the IR, i.e. $m_{\textrm{eff}}^2=6$. If one just uses this lower bulk mass over the full range of $r$, then the predicted baryon mass falls to $M_B=1.5$ GeV.\\

This naive result certainly suggests that the anomalous dimension of a quark operator can play a role in reducing the predicted mass of its associated bound state. In the following section, we seek a more accurate estimate by generating the IR scale dynamically via a running $\gamma$.

\section{Dynamic AdS/QCD Model}

A more advanced AdS/QCD model would need to incorporate the running anomalous dimension $\gamma$ of the quark anti-quark bilinear, which triggers a BF bound violation and dynamically generates the mass gap, as well as the running dimensions of all the operators that source our bound states. We will work in the spirit of the Dynamic AdS/QCD model outlined in \cite{Alho:2013dka}. The model, analogous to some of the earliest hard wall models, attempts to make the minimum number of changes to a basic AdS/QCD model that will exemplify the physics under discussion without it being a first principle holographic description of QCD. Mostly, we import insights from the D3/probe D7 system \cite{Karch:2002sh, Kruczenski:2003be, Erdmenger:2007cm}.

\subsection{Anomalous Dimension for the Quark Condensate}
 To describe the quark condensate dynamics, we allow ourselves to be guided by the D3/probe D7 system, where the $\bar{q} q$ operator is dual to a scalar field $L$ of dimension one (i.e. $L=r \chi$) in AdS space - the field $\chi$ has bulk mass $m^2 = -3$ ($\Delta = 3)$. By considering the  action
\begin{equation}
    S = \int d^4x dr ~\left (r^3 (\partial_r L)^2 + r \Delta m^2 L^2 \right) \label{eq:scalar_action}
\end{equation}
one can see that, upon rewriting the first term in terms of $\chi$ and integrating by parts, we arrive at the canonical Lagrangian for the dimensionless AdS field $\chi$ with $m^2 = -3 + \Delta m^2$. Since the BF bound is  $m^2 \geq -4$ in the $AdS_5$ spacetime, it is clear that a scalar mass deviation of $\Delta m^2 < -1$ will trigger the violation we are looking for.\\

The equation of motion associated with \eqref{eq:scalar_action} is given by
\begin{equation}
    \partial_r\left[r^3 \partial_r L \right]  + r \Delta m^2  L= 0 \label{eq:scalar_eom} 
\end{equation}
Substituting (at the level of the equation of motion)

\begin{figure}[h!]
    \centering
    \includegraphics[scale=0.3]{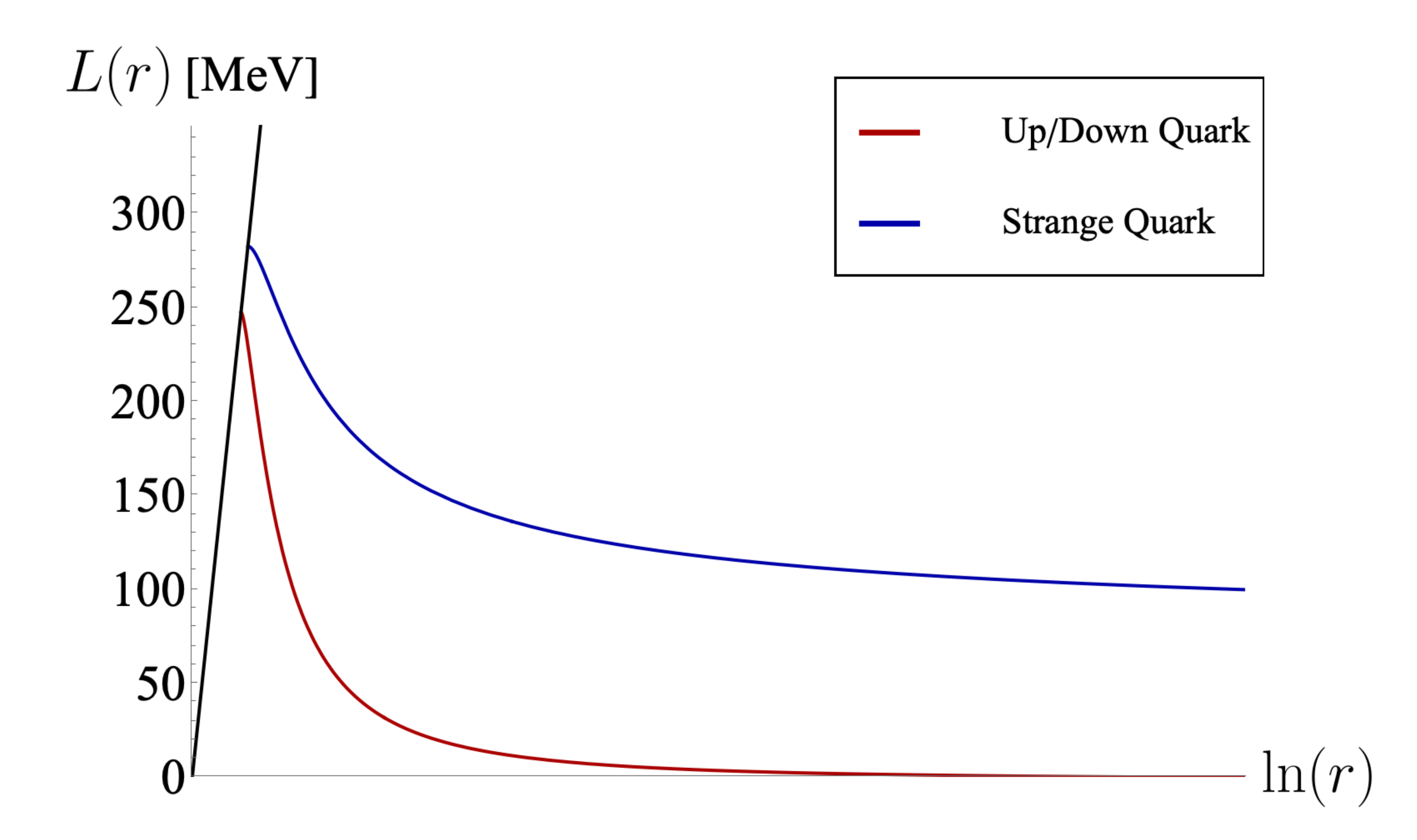}
    \caption{Plot of the bulk scalars $L_{ud}$ and $L_s$ as functions of the RG scale $r$, with the black line indicating the on-mass-shell condition. In the UV, these fields take on the values of the up/down quark mass (set to zero) and the strange mass (100 MeV) respectively. At lower energies, a quark condensate forms, contributing to the mass of each scalar in the IR.}
    \label{fig:brane_embeddings}
\end{figure}

$\Delta= 3 + \gamma$ and expanding $m^2 = \Delta(\Delta-4)$ for small $\gamma$ then gives
\begin{equation}
    \Delta m^2 = 2 \gamma = -\frac{4}{\pi} \alpha
\end{equation}
i.e. a scalar bulk mass that depends on $r$ is dual to the running anomalous dimension of its associated boundary operator. Using \eqref{eq:op_dim_general}, we then express $\gamma$ in terms of $\alpha$, which is specified by the following RG flow at one loop
\begin{equation}
    \mu { d \over d \mu} \alpha = -b_0 \alpha^2, \hspace{1cm} b_0 = {1 \over 6 \pi} \left(11 C_2(G) -4 N_f T(F) \right)
\end{equation}

For an $SU(N_c)$ gauge theory like QCD, the quadratic Casimir for the adjoint representation specifies the number of colours, i.e. $C_2(G)=N_c$. In the fundamental representation, the trace over group generators is normalised by $T(F)=1/2$.

Naively, one would assume $\alpha$ is a function of $r$ only, but this would remove any stable solutions below where the BF bound is violated. The natural resolution (found in the D3/probe D7 system) is to set $r^2 \rightarrow r^2 + L^2$, which is both dimensionally sensible and means that, if $L$ grows sufficiently, the BF bound violation is removed and a stable solution is allowed to form.

To solve \eqref{eq:scalar_eom} for the vacuum configuration of the theory, we must impose sensible boundary conditions on $L$ in the IR region. We use the regularity condition $\partial_\rho L=0$ proposed by the D3/probe D7 system, thus allowing $L$ to be interpreted as the RG flow of the constituent mass.

We should also integrate any quarks from the theory which fail to satisfy the on-mass-shell condition $L \leq r$. As such, we begin the solution on the line $L=r$ with $\partial_\rho L=0$ and only explore scales above that bound. We show vacuum solutions that asymptote to either zero quark mass or the strange mass in the UV in fig.\ref{fig:brane_embeddings}.

Goldstone bosons, associated with the broken axial symmetry, correspond to the phase of the field $L$. As usual in the D3/probe D7 system \cite{Kruczenski:2003uq}, the pion wave functions $\pi(r)$ are described precisely by the vacuum $L$ solutions above (for zero UV quark mass). The additional interpretation of these solutions does, however, require that they be properly normalised in order to have a canonically kinetic term in the 4D Lagrangian. As such, we impose the following normalisation condition on the pion wave function

\begin{equation}
    \int dr { r^3\over (r^2 + L^2)^2} \pi(r)^2 = \frac{1}{2} \label{eq:pion_normalisation}
\end{equation}

\subsection{Nucleon Mass Revisited}
In order to calculate the spin-half baryon mass in the dynamical model, we must first recalculate the scalar mass of the $\rho$ meson ground state. The equation of motion that determines the $\rho$ meson mass is now given by
\begin{equation}
    \partial_r r^3 \partial_r A + {r^3 M^2 \over (r^2+L^2)^2}A = 0 \label{eq:vector_eom}
\end{equation}
We solve \eqref{eq:vector_eom} with $A(r_{\rm min})=1$ and $A'(r_{\rm min})=0$ with $r_{\rm min}$ specifying where the quark mass goes on-shell. We find a new value for the bulk mass of the $\rho$ meson that will set the QCD scale.

As for the nucleon, the equation of motion for a bulk fermion in the Dynamic AdS/QCD model was computed in \cite{Abt:2019tas}. The nucleon wave function $\psi$ is given by the solution of
\begin{equation} \begin{array}{c}
\left( \partial_r^2 + \mathcal{P}_{1} \partial_r + \frac{M^2_{B}}{(r^2+L^2)^2} + \mathcal{P}_2 \frac{1}{(r^2+L^2)^2} \right. \\ \\\left. \hspace{1cm} -  \frac{m_\psi^2}{(r^2+L^2)} + \mathcal{P}_3 \frac{m_\psi}{(r^2+L^2)^{3/2}}\right) \psi = 0 \end{array} \label{eq:eom_fermions} 
\end{equation}
where $M_B$ is the baryon mass, $m_\psi = 5/2$ as before and the pre-factors are given by
\begin{align*}
    \mathcal{P}_1 &= \frac{6}{r^2+L^2} \left( r + L~ \partial_{r}L \right), \\
    \mathcal{P}_2 &= 2 \left( (r^2 + L^2) L \partial^2_{r} L + (r^2 + 3L^2) (\partial_{r}L)^2 + 4 r L \partial_{r}L \right. \\ & \left. \hspace{2cm} + 3 r^2 + L^2  \right), \\
    \mathcal{P}_3 &= \left( r+ L ~ \partial_{r}L   \right).
\end{align*}

We can add a contribution to the running anomalous dimension of the three quark operator by allowing the fermion mass to be $r$ dependent. Using the relation between $m_\psi$ and the dimension of the operator $\Delta$, as well as the one loop running result for a three quark operator, we find that
\begin{equation}
    \Delta m_\psi = \gamma = -{3 \over \pi} \alpha \label{eq:anom_dim_baryon}
\end{equation}

The normalisation of the nucleon wave function is
\begin{equation}
    \int dr { r^3\over (r^2 + L^2)^{1/2}} \psi(r)^2 = \frac{1}{2}
\end{equation}

Amending $m_\psi$ with the running anomalous dimension \eqref{eq:anom_dim_baryon}, we can solve \eqref{eq:eom_fermions} subject to the usual IR boundary conditions. Then, tuning $M_B$ such that $\psi$ vanishes in the UV (the normalised wave function is plotted in fig.\ref{fig:norm_wavefnc}), we find that $M_B = 1.40 M_\rho = 1.08$ GeV, which is within 15\% of the measured proton mass ($938$ MeV).

This result is smaller than we achieved in the hard wall model in which the dimension of the three quark operator was fixed at $\Delta = 3$. This is because the value of the anomalous dimension $\gamma$ is greater in this model to the extent that it violates the BF bound very close to the chiral symmetry breaking scale. Such a violation within a short region of $r$ need not be sufficient to trigger an instability. Instead, the contribution provided by the `kinetic' derivative terms in the $r$ direction may counter mass terms in the field potential, leading to an overall solution that remains stable.

With the stability of the solution assumed, the phenomenological effect of the running dimension is to drive the mass of the bound state down. The fact that the one loop result for $\gamma$ reproduces a nucleon mass that is closer to the physical value suggests that a large running of the anomalous dimension may indeed be present. Of course, one could tune $\gamma$ in the non-perturbative regime in a way that reproduces the observed mass exactly, though without an analytical understanding of this phenomena it would be difficult to motivate any kind of precise ansatz.

\subsection{Strange Baryon Masses}
We next turn our attention to QCD bound states containing strange quarks, such as the $\Lambda$ baryon. For states containing yet heavier quarks, we expect that the running anomalous dimension has a negligible impact on the bound state masses. This is because the dimension of the quark operator will only run as far as its mass scale which, for those heavier than the strange quark, is large enough to suppress the effects the light quarks see from the sudden pole in the QCD coupling.

In top-down models with flavour branes, the strange and light quark branes will separate in the bulk space and mixed heavy-light states would appear as complicated stringy states tied between them \cite{Erdmenger:2007vj}. We do not try to reproduce this structure here. The key point is that the bound state operators see the $L(r)$ functions associated with the quarks they contain, thus making them aware of the constituent's masses. A simple phenomenological approach is to write
\begin{equation}
    L^2 \rightarrow f_{ud} L_{ud}^2 + f_s L_s^2 \label{eq:one_strange_structure}
\end{equation} 

where $f_i$ is the fraction of the quarks of the type $i$ in the hadron.

For example, we consider a \textit{uds} $\Lambda$ bound state whose equation of motion follows the same general form as that of the nucleon \eqref{eq:eom_fermions} but with an amended field structure $L^2 \rightarrow \frac{2}{3}L_{ud}^2 + \frac{1}{3} L_s^2$ reflecting the composition of the $\Lambda$ baryon in terms of first and second generational quarks. The distinction made is that, for the scalar dual to the strange quark $L_s$, the solution asymptotes to a non-zero bulk mass in the UV limit (fig.\ref{fig:brane_embeddings}). This field redefinition changes both the pre-factors of the equations of motion and the deformed AdS radius, which is now given by $(r^2 + \frac{2}{3}L_{ud}^2 + \frac{1}{3} L_s^2)^{1/2}$.

Solving the equation of motion for the $\Lambda$ returns a wave function solution $\psi_\Lambda$ which vanishes in the UV limit for a tuned baryon mass of $M_\Lambda = 1.49 M_\rho = 1.15$ GeV (fig.\ref{fig:norm_wavefnc}). This prediction is within 3\% of the measured $\Lambda$ mass ($1115$ MeV).

Moving to heavier bound states, we can once again amend the field structure and deformed AdS radius to describe particles comprised of two strange quarks, such as $uss$. In this case, we take \eqref{eq:one_strange_structure} with $f_{ud} = 1/3$ and $f_{s}=2/3$. Solving a similar equation of motion as with the $\Lambda$, we find a $\Xi$ mass of 1.22 GeV, which is within 8\% of its measured value (1315 MeV). Again, the running anomalous dimension of the quark fields is key to producing these closer fits with experimental data.

\section{Sexaquarks in AdS/QCD}

Given the apparent importance of running anomalous dimensions in the baryonic sector, it is interesting to look for other light quark states in which the same mechanism might be present. Looking towards novel forms of matter, it has been suggested that the spectrum of QCD may include a deeply bound six-quark state (see for example \cite{Farrar:2017eqq, Farrar:2022mih, Azizi:2019xla} to motivate this and review the phenomenology). The $uuddss$ state forms a colour singlet that is antisymmetric in colour, flavour and spin, endowing it with a spatially symmetric wave function. Where the state is considered as a loosely bound state of two $\Lambda$ baryons, the $uuddss$ state has historically been referred to as the H dibaryon \cite{Jaffe:1976yi}.\\

Naively, one might estimate the H dibaryon mass to be three times that of the rho meson plus an additional contribution of 200 MeV arising from the larger mass of the two strange quarks. This leads to a mass of around 2.5 GeV. A previous bag model analysis placed this estimate a little lower at 2.15 GeV \cite{Jaffe:1976yi}. Both these predictions lie above the $\sim$2 GeV stability bound which causes the dibaryon to decay with a typical weak interaction lifetime of $10^{-10}\textrm{s}$. Lattice QCD simulations \cite{NPLQCD:2012mex,PadmanathMadanagopalan:2021exb} are currently unable to work at the physical quark masses. One could consider models with larger quark masses, but this may exclude the key regime around the scale of chiral symmetry breaking thus impacting the mass predictions.

At present, experimental searches for the dibaryon remain unsuccessful. If the state were more tightly bound, its mass could sit below 2 GeV, causing it to potentially remain undetected against the large neutron background in collider experiments \cite{Farrar:2017eqq}. If confirmed, it is proposed that this tightly bound \textit{sexaquark} would not only be stable, but it might also be a good candidate for dark matter \cite{Farrar:2018hac} or a stable cosmological relic \cite{Kolb:2018bxv}. 

We take this discussion as motivation to investigate the sexaquark state in holography. 
A previous estimate of the sexaquark mass within holography was conducted in the Sakai-Sugimoto model \cite{Suganuma:2016lmp}, where the sexaquark was treated as an instanton-like solution arising as a bound state of baryons - that study predicted a mass of 1.7 GeV. Instead, we will model the sexaquark as a scalar in the bulk of the Dynamic AdS/QCD model. If there is a deeply bound state, this may be the more appropriate description, distinct from treating it as a bound state of two $\Lambda$s.  The scalar's bulk mass will run in our description, dual to the anomalous dimension of the $uuddss$ operator.

\subsection{Sexaquarks in the Hard Wall Model}
We can represent the sexaquark operator by inserting a scalar field $\Phi$ into $AdS_5$ bulk space with Klein-Gordon action
\begin{equation}
    S = \int d^5x \sqrt{-\textrm{det}g} \left((\partial_\mu \Phi)(\partial_\nu \Phi^*) g^{\mu\nu} + m^2\Phi^2\right)
\end{equation}
Writing the field as a plane wave $\Phi(r,x) = \phi(r) e^{ikx}$ with $k^2 = -M^2$, the resulting equation of motion
\begin{equation}
    \partial_r\left[r^5 \partial_r \phi \right] - r^3m^2\phi + M^2 r \phi= 0 \label{eq:eom_r_only}
\end{equation}
admits the following large $r$ solution for $m^2 = \Delta(\Delta - 4)$
\begin{equation}
    \phi(r) = \frac{C_1}{r^{\Delta}} + \frac{C_2}{r^{4-\Delta}}
\end{equation}
Using the UV scaling dimension of the six quark state $\Delta = 9$, we conclude that the bulk mass $m^2 = 45$.

Taking the IR boundary condition $\phi'({\rm wall})=0$, $M^2$ is again tuned such that $\phi$ vanishes in the UV, leading to a sexaquark mass of $M = 3.77 M_\rho \approx 2.90$ GeV. This large estimate assumes that the dimension of the quark is fixed at its maximum value throughout the strong coupling regime. Alternatively, one could assume that the IR dimension $\Delta = 6$ (as suggested by the one loop calculation) holds for all energies instead. In this case, the sexaquark mass is substantially lower at 1.90 GeV.

With two very distinct estimates for the sexaquark mass residing on either side of the 2 GeV stability condition, this simple hard wall model demonstrates the importance that the specific form of the running dimension likely has on the calculation of bound state masses composed of a greater number of quarks. As such, a prediction of the sexaquark mass in Dynamic AdS/QCD may be particularly worthwhile.

\subsection{Dynamic AdS/QCD and the Sexaquark}
We will treat the sexaquark operator as a scalar in the spirit of  \eqref{eq:scalar_action} but with a non-zero spatial derivative term
\begin{equation}
    S = \int d^4x dr ~ r^3 \left[(\partial_r L)^2 + \frac{(\partial_x L)^2}{(r^2 + L^2)^2} + \frac{\Delta m^2 L^2
}{r^2}\right]
\end{equation}
Assuming a plane wave solution and adjusting the factor of $L^2$ to average over the quark constituents, we arrive at the following equation of motion,
\begin{equation}
    \partial_r\left[r^3 \partial_r L \right]  - r \Delta m^2  L + {r^3 M^2 \over (r^2 +{2 \over 3}L_{ud}^2 +{1 \over 3} L_s^2)^2} L = 0 \label{eq:sex_eom}
\end{equation}
where $M^2$ is the sexaquark mass. To enforce a UV bulk mass of $m^2=45$ and the IR running from the one loop analysis, we set
\begin{equation}
    \Delta m^2 = (3+ m^2) + 14\gamma = 48 - {84 \over \pi} \alpha
\end{equation}
where we have used \eqref{eq:op_dim_general} with $n=6$.

As in the previous cases, we now seek a solution to \eqref{eq:sex_eom} for which the UV field vanishes. We find that this is satisfied for $M_S = 2.27 M_\rho \approx 1.75$ GeV. This result would support the stable deeply bound hypothesis, although it is clearly very dependent on the assumptions made about the running dimension of the six-quark operator in the

\begin{figure}[h!]
    \centering
    \includegraphics[scale=0.3]{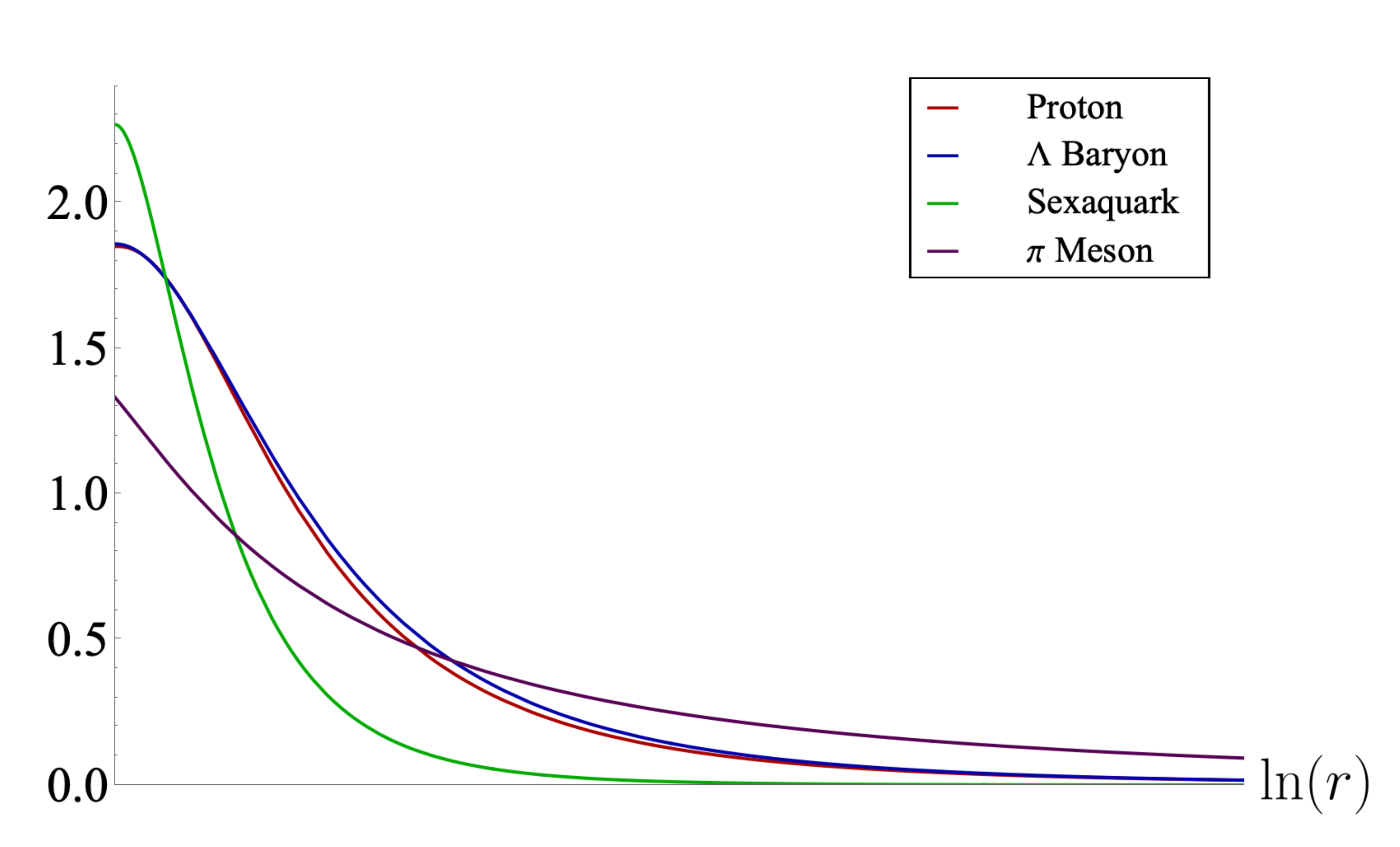}
    \caption{Plot of the normalised wave functions for the proton, $\Lambda$ baryon, sexaquark and $\pi$ meson as a function of RG scale $r$. For each wave function, the value of the bound state mass has been tuned such that the solution vanishes in the far UV.}
    \label{fig:norm_wavefnc}
\end{figure}

IR. To stress that point, one could perform a similar analysis for a di-neutron state and get an equally low mass, which appears to be at odds with experimental data. Whether the unique wave function of the sexaquark would enter beyond one loop to justify the assumptions here is unclear. Nevertheless, holography can entertain the possibility that the sexaquark is stable.

Similarly to the pion wave function \eqref{eq:pion_normalisation}, the normalisation of the sexaquark wave function is fixed by
\begin{equation}
    \int dr { r^3\over (r^2 +{2 \over 3}L_{ud}^2 +{1 \over 3} L_s^2)^2} L(r)^2 = \frac{1}{2}
\end{equation}

We plot the normalised wave functions for the nucleon, $\Lambda$, sexaquark and pion in fig.\ref{fig:norm_wavefnc}. We note that any interactions between these states will be set by overlap integrals of these wave functions; the ratios of which all appear to be of order 1. One should also not expect any large factors from the calculation of these integrals which act to raise or lower particular interactions. A priori, the sexaquark does not seem particularly decoupled from the rest of the hadronic spectrum. However, it is important to stress that these calculations, whose origin is in the large $N_c$ limit, do not include secondary interactions such as the size of pion clouds surrounding each state.

\vspace{3em}
\section{Discussion}

In this paper we have employed some simple holographic models to explore the importance of the running anomalous dimension of operators on the predicted spectrum of bound state masses in QCD. Similar considerations have previously been made for the sigma meson in models that dynamically generate chiral symmetry breaking. In those models, the running dimension drives the sigma to become tachyonic, triggering the chiral symmetry breaking. This suggests that the running for other light quark operators might be important. 

Our first investigation was of the nucleon mass. In hard wall models using the UV dimension of the three-quark operator, the nucleon mass is typically high relative to the mass of the rho meson. We instead enacted a model that includes the running of the anomalous dimension for the quark condensate and dynamically describes chiral symmetry breaking in place of the hard wall. Using an ansatz for the bulk fermion, we found that the nucleon mass was brought down to 1.1 GeV. Our ansatz matches the bulk mass to the one loop QCD running of the operator dimension, but then extrapolates that beyond the UV regime in an uncontrolled fashion. Even so, we conclude that the running can play a significant role in the predicted nucleon mass and should not be neglected. A similar analysis for the $\Lambda$ ($usd$) and $\Xi$ ($uss$) baryons also brings their masses into line with experimental values at the 3 and 8 percent level respectively.

We also considered a more exotic singlet sexaquark state ($uuddss$). Including the one loop running of $\gamma$ in this case generated a mass prediction of 1.75 GeV, a value similar to other holographic models and low enough to make the state stable against decay into two $\Lambda$ baryons. Whilst the precise value depends on the assumed running, such a stable state does not seem incompatible with holography.

Overall, we conclude that the running anomalous dimensions of multi-quark operators is important when determining the bound state spectrum and holographic models should seek to include these in the future.

\bigskip \noindent {\bf Acknowledgements:}  
NE's work was supported by the
STFC consolidated grant ST/T000775/1.


\end{document}